\documentclass[aps,twocolumn,preprintnumbers,showpacs]{revtex4-1}
\usepackage{amssymb}
\usepackage{mathrsfs}
\usepackage{float}
\usepackage{bm}
\usepackage{amsmath}
\usepackage{graphicx}
\usepackage{subfigure}
\usepackage{amsmath}
\usepackage{array}
\usepackage{lipsum}
\begin{document}
\newcommand*{\PKU}{School of Physics, Peking University, Beijing 100871,
China}\affiliation{\PKU}

\title{Final state interactions at the threshold of Higgs boson pair production}

\author{Zhentao Zhang}\email{zhangzt@pku.edu.cn}\affiliation{\PKU}

\begin{abstract}
We study the effect of final state interactions at the threshold of Higgs boson pair production in the Glashow-Weinberg-Salam model. We consider three major processes of the pair production in the model: lepton pair annihilation, ZZ fusion, and WW fusion. We find that the corrections caused by the effect for these processes are markedly different. According to our results, the effect can cause non-negligible corrections to the cross sections for lepton pair annihilation and small corrections for ZZ fusion, and this effect is negligible for WW fusion.
\end{abstract}
\maketitle

Half a century after the construction of the Glashow-Weinberg-Salam~(GWS)~model of electroweak interactions, the observation of Higgs boson at the Large Hadron Collider~(LHC) is its latest triumph \cite{ATLAS,CMS}. In the future, further determination of the properties of Higgs boson will be extremely important for us to understand the fundamental laws of nature. Any deviation from the predictions in the Standard Model can give us a valuable clue to the long-hunted new physics. In order to ascertain the role of Higgs field in the Standard Model, we need to measure its couplings to fermions and gauge bosons. Furthermore, to reconstruct the details of the Higgs potential, we have to precisely determine the Higgs self-couplings.

At present, an important task of us is to explore potential effects for the Higgs self-interactions, and the Brout-Englert-Higgs mechanism must be precisely tested from low to high energies. In this paper, we shall investigate the effect of final state interactions at the threshold of Higgs boson pair production in the GWS model. This effect is tied to the low-energy properties of the Higgs self-interactions.

After the electroweak symmetry breaking, the Higgs self-interactions can be written in the form
\begin{equation}
V(H)=\frac{1}{2}{m^2_H}{H}^2+\lambda{v}{H}^3+\frac{\lambda}{4}H^4.
\end{equation}
For the mass of Higgs boson $m_H\simeq125.5~\text{GeV}$ and the vacuum expectation value $v\simeq246~\text{GeV}$,  $\lambda=m^{2}_{H}/2v^2\approx0.130$.

\begin{figure}[h]
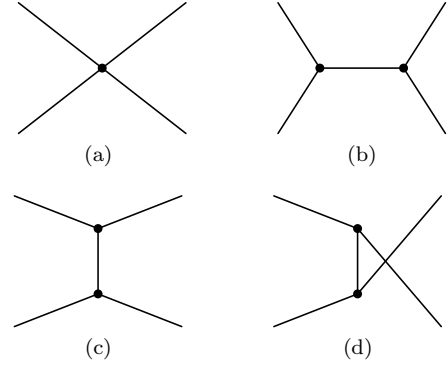

\center
 \subfigure[]{ \includegraphics[width=2.8cm]{fig1a.1}}~~~~~
 \subfigure[]{\includegraphics[width=2.8cm]{fig1b.1}}\\
 \subfigure[]{\includegraphics[width=2.8cm]{fig1c.1}}~~~~~
 \subfigure[]{\includegraphics[width=2.8cm]{fig1d.1}}
   \caption{The tree-level diagrams for the scattering of two Higgs bosons in the Standard Model. Solid line denotes Higgs boson.}\label{1}

\end{figure}

To establish the non-relativistic potential for the Higgs self-interactions, we need to consider the scattering process $HH\rightarrow HH$ to leading order in $\lambda$. There are four Feynman diagrams that contribute, see Fig.~\ref{1}.

Note that we can separately discuss the contributions from each diagram, since there is Bose statistics involved in the process. In the beginning, let us consider the contributions from the quadrilinear self-coupling of Higgs boson. The diagram is shown in Fig.~\ref{1}a, and its amplitude is
\begin{equation}
  i\mathcal{M}=-i6\lambda.\label{four}
\end{equation}
According to the definition of Born scattering amplitude in non-relativistic quantum mechanics, the non-relativistic potential in the momentum spaces is
\begin{equation}
  \widetilde{V}_{\text{quad}}(\bm{q})=\frac{3\lambda}{2m^2_H}.
\end{equation}
Using the Fourier transform to $\widetilde{V}(\bm{q})$, we can get the non-relativistic potential for this term
\begin{equation}\label{quad}
V_{\text{quad}}(\bm{r})=\frac{3\lambda}{2m^2_H}\delta^{(3)}(\bm{r}).
\end{equation}
The presence of the delta function comes from the specially local structure in the diagram.

The amplitude for the ``annihilation'' diagram in Fig.~\ref{1}b is
\begin{equation}
 i\mathcal{M}=-18\lambda m^2_H\frac{i}{(p_1+p_2)^2-m^2_H},
\end{equation}
where $p_1$ and $p_2$ are the 4-momenta of the incoming Higgs bosons. The amplitude in the non-relativistic domain can be approximated as Eq.~(\ref{four}), and then we can get a well defined non-relativistic potential. It is not necessary to repeat the same calculations. The non-relativistic potential for the ``annihilation'' diagram is defined as
\begin{equation}\label{quad}
V_{\text{ann}}(\bm{r})=\frac{3\lambda}{2m^2_H}\delta^{(3)}(\bm{r}).
\end{equation}

The amplitude for the ``exchange'' diagram in Fig.~\ref{1}c is
\begin{equation}
 i\mathcal{M}=-18\lambda m^2_H\frac{i}{q^2-m^2_H},
\end{equation}
where 4-momentum q is the momentum transfer. In the non-relativistic domain, the amplitude becomes
\begin{equation}
 i\mathcal{M}=18\lambda m^2_H\frac{i}{\mid\bm{q}\mid^2+m^2_H},
\end{equation}
and the non-relativistic potential in momentum space is
\begin{equation}
  \widetilde{V}_{\text{ex}}(\bm{q})=-\frac{9\lambda}{2}\frac{1}{\mid\bm{q}\mid^2+m^2_H}.
\end{equation}
After the Fourier transform, the non-relativistic potential for the ``exchange'' term is
\begin{equation}\label{ex}
 V_{\text{ex}}(\bm{r})=-\frac{9\lambda}{8\pi r}e^{-m_Hr}.
\end{equation}

The presence of the ``cross'' term in Fig.~\ref{1}d comes from the indistinguishable property of the identical particles. However, the ``cross'' term for the scattering of two identical particles does not contribute to the non-relativistic potential. It is because that the principle of identity also needs to be considered in the scattering theory of non-relativistic quantum mechanics, and then the non-relativistic potential that we get from the elastic scattering of the distinguishable particles will automatically include the contributions from the ``cross'' term of the identical particles. In appendix, we shall show an elementary example in detail.

As a result, we obtain the non-relativistic potential for the Higgs self-interactions
\begin{align}\label{all}
  V(\bm{r})= &V_{\text{quad}}(\bm{r})+V_{\text{ann}}(\bm{r})+V_{\text{ex}}(\bm{r})\nonumber\\
           =&\frac{3\lambda}{m^2_H}\delta^{(3)}(\bm{r})-\frac{\alpha}{ r}e^{-m_Hr},
\end{align}
where coupling constant $\alpha=9\lambda/(8\pi)$.

We know that the probability of scattering is in proportion to the squared modulus of the wave function of the created particles and a non-relativistic potential in the final state can alter the amplitude of the final state wave function in the reaction zone~\cite{Landau}. Notice that Higgs boson can be treated as a stable particle for the Higgs self-interactions, since the width of Higgs boson is only a few MeV \cite{PDG}. Therefore, the non-relativistic Higgs potential in the final state might be able to modify the relevant cross section in perturbation theory, and this effect can be very important for us to understand the low-energy properties of the Higgs self-interactions. After getting potential (\ref{all}), we shall investigate the effect of final state interactions at the threshold of Higgs boson pair production. In recent years a similar effect, which considering the distortion of the incoming wave function of the initial particles, in dark matter was discussed, see. e.g.,~\cite{Hisano,Cirelli,Arkani-Hamed,Lattanzi,Iengo,Zhang}.

\begin{figure}[h]
\center
\subfigure[]{\includegraphics[width=3.4cm]{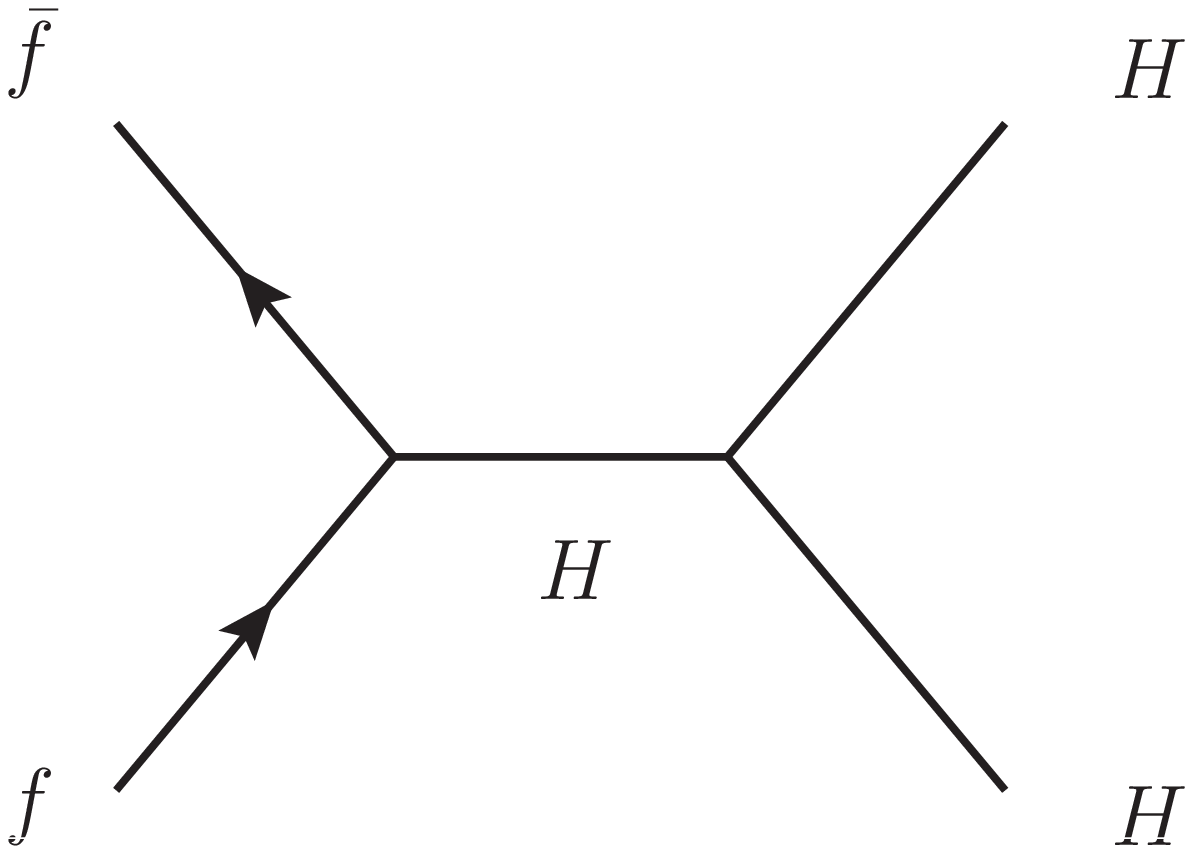}}~~~~
\subfigure[]{\includegraphics[width=3.4cm]{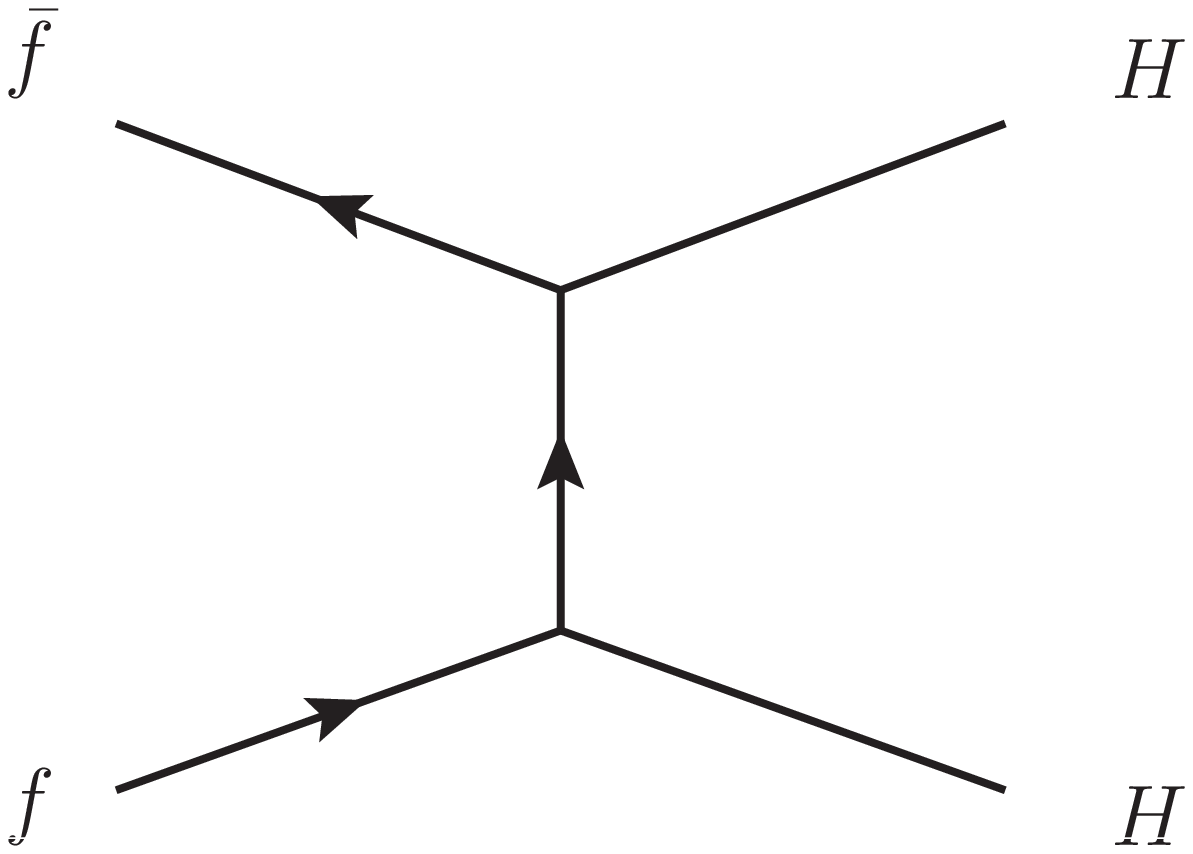}}
\subfigure[]{\includegraphics[width=3.4cm]{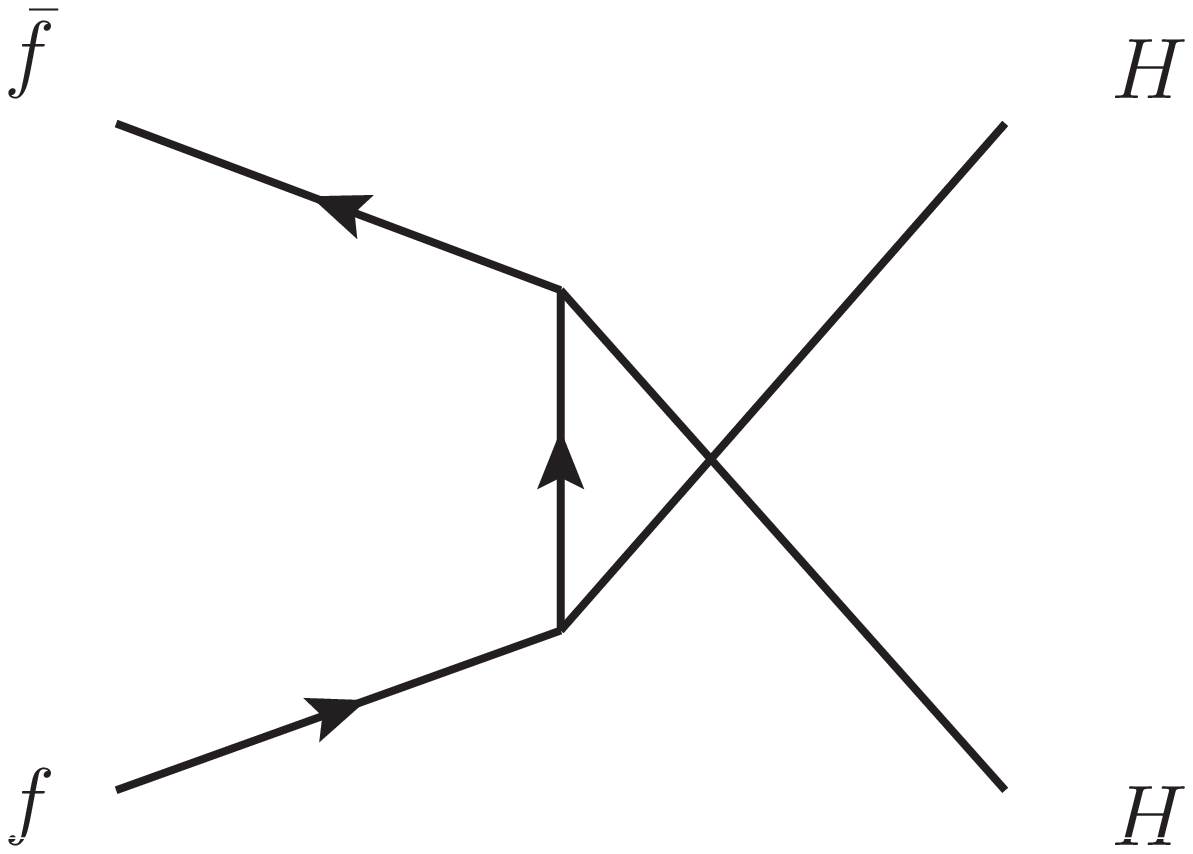}}~~~~
\subfigure[]{\includegraphics[width=3.4cm]{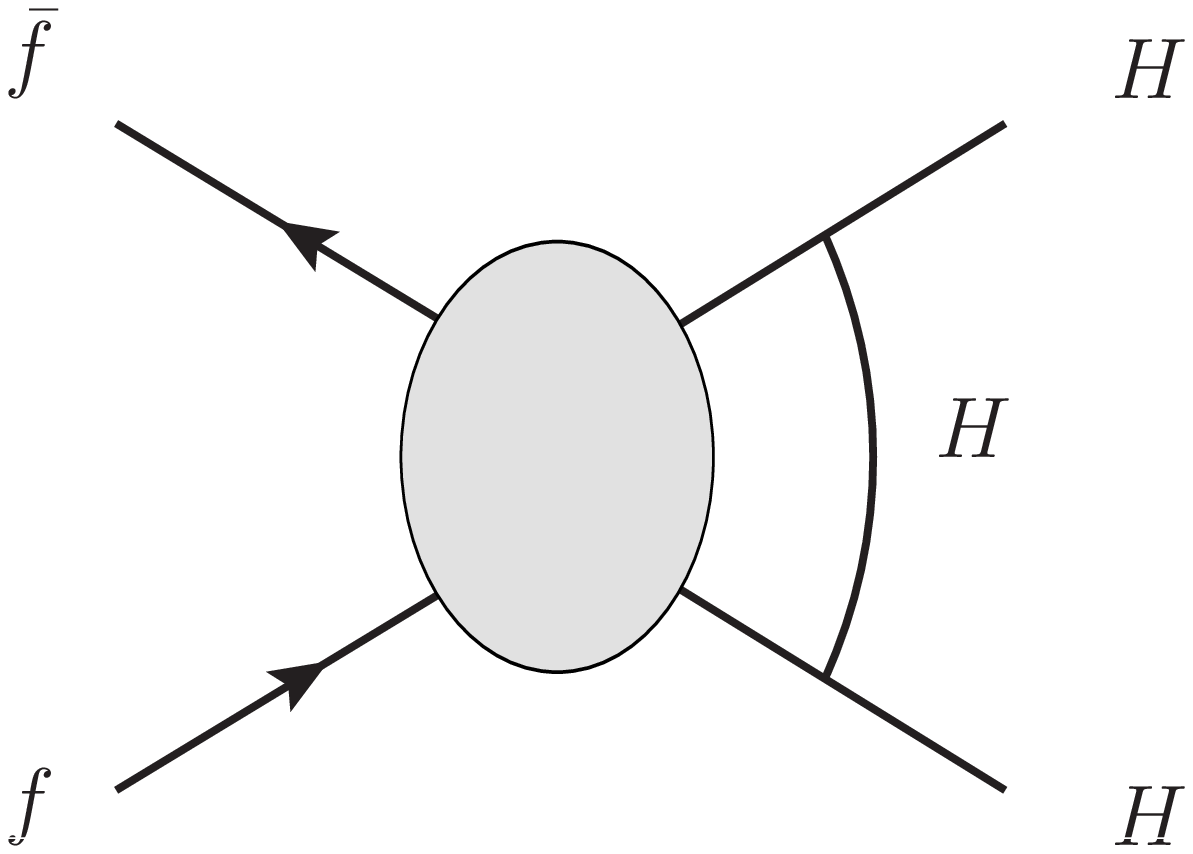}}
\caption{Higgs boson pair production in the processes of lepton pair annihilation (a-c), and the generic diagram of the Higgs self-interactions in the final state (d).}\label{pair}

\end{figure}

At first let us consider the final state interactions in the processes of lepton pair annihilation ${f\overline{f}\rightarrow HH}$, see Fig.~\ref{pair}. We know that in perturbation theory the scattering amplitude is defined in the plane wave bases. However, as we mentioned above, due to the appearance of the interactions in the non-relativistic region of the processes, the wave functions in the reaction zone can be very different from the plane waves. To consider the corrections in the processes, the universal amplitudes for the $s$-, $t$-, and $u$-channel diagrams in Fig.~(\ref{pair}) can be written in the forms
\begin{align}
\label{uni1}
\mathcal{M}_{s}&=\mathcal{M}^{p}_{s}\psi_i(0)\psi^{*}_f(0)/\psi^{0}_i(0)\psi^{0*}_f(0),\\
\label{uni2}
\mathcal{M}_{t}&=\mathcal{M}^{p}_{t}\psi_i({\bm r_t})\psi^{*}_f({\bm r_t})/\psi^{0}_i({\bm r_t})\psi^{0*}_f({\bm r_t}),\\
\label{uni3}
\mathcal{M}_{u}&=\mathcal{M}^{p}_{u}\psi_i({\bm r_u})\psi^{*}_f({\bm r_u})/\psi^{0}_i({\bm r_u})\psi^{0*}_f({\bm r_u}),
\end{align}
where $\psi_i$ and $\psi_f$ are the wave functions for the relative motions of the particles in the initial and final systems, $\psi^{0}_i=e^{i{\bm{k_i\cdot r}}}$ and $\psi^{0}_f=e^{i{\bm{k_f\cdot r}}}$ are the plane waves for free motions in the initial and final systems, ${\bm r_{t}}$ and ${\bm r_{u}}$ denote the separate distances of the two vertexes in $t$- and $u$-channel diagrams, and $\mathcal{M}^{p}$ denotes the relevant amplitude in perturbation theory.

The wave functions introduced in universal amplitudes (\ref{uni1}-\ref{uni3}) can give the corrections to the squared modulus of the plane waves in the reaction zone, and we can solve the Schr\"{o}dinger equations to get the wave functions if we know the forces between the particles in the initial and final states. Notice that in the perturbation region the universal amplitudes become the same as the amplitudes in perturbation theory.

In the process of the electron-positron annihilation, it is straightforward to see that the contributions of $t$- and $u$-channel diagrams in Fig.~(\ref{pair}) can be ignored, since the Yukawa coupling to electron ($m_e/v$) is very weaker than the Higgs self-couplings. Therefore, in practical applications, the squared modulus of the amplitude at the threshold can be simplified as
\begin{equation}\label{ep}
 |\mathcal{M}|^2_{e\overline{e}\rightarrow HH}=|\mathcal{M}^{p}_{s}|^2_{e\overline{e}\rightarrow HH}|\psi^{*}_f(0)|^2,
\end{equation}
where $|\psi^{*}_f(0)|^2$ is the correction term caused by the final state interactions in the process.

Note that to find the corrections caused by the effect of final state interactions, we only need to consider the corrections to the s-wave, because the Higgs boson pair is emitted in an s-wave at the threshold. The reduced two-body radical Schrodinger equation with a potential $V({r})$ is ${d^2}\phi(r)/d{r^2} - mV({r})\phi(r) =-{(mv)^2}\phi(r)$, where $\phi(r)=rR_0(r)$, and $R_0$ is the s-wave radical function. The radical wave function $R_{kl}$ can be normalized as $\int^{\infty}_{0}r^2R_{k'l}R_{kl}dr=2\pi\delta(k'-k)$, and notice that $R_{kl}$ is real in this convention. Since we have already obtained non-relativistic potential~(\ref{all}), to find $|\psi^{*}_f(0)|^2$ in Eq.~(\ref{ep}), we can numerically solve the Schr\"{o}dinger equation with boundary conditions $\phi(r)\rightarrow 0$ as $r\rightarrow 0$ and $\phi(r)\rightarrow 2\sin{(kr+\delta_0)}$ as $r\rightarrow \infty$. However, we should note here that although the non-relativistic potential for the Higgs self-interactions is composed of different terms, only the ``exchange'' term would be involved in the effect. The reason is that the final state interactions between two separate particles come from the exchange of their force carriers, thus only the ``exchange'' part of the non-relativistic potential need be considering. The numeric simulations for the 125.5 GeV Higgs boson are presented in Tab.~\ref{electron}. We find that at the threshold the two-Higgs-boson final state interactions can increase the cross section by near ten percent.

\begin{table}[h]
\center
\caption{The numeric simulations for the corrections in electron-positron annihilation process, where $\sqrt{s}$ is the center of mass energy.}
\begin{tabular}{cc} \hline \hline

$~~~~~~~~~~~~~\sqrt{s}$(GeV)~~~~~~~~~~~~~~~~ &~~~~~~~~~~Correction(\%)~~~~~~~ \\\hline
251.1~~&9.84\\
251.2~~&9.83 \\
251.3~~&9.82 \\ \hline\hline
\end{tabular}
\label{electron}
\end{table}

The strength of the effect grows slowly close to the threshold, and it is interesting to investigate that how the largest corrections can be reached. Using the principle of detailed balancing~\cite{Landau}
 \begin{equation}\label{pdb}
  \frac{1}{p^2_{f}}\frac{d\sigma}{d\Omega}({i\rightarrow f})= \frac{1}{p^2_{i}}\frac{d\sigma}{d\Omega}({f^{*}\rightarrow i^{*}}),
 \end{equation}
where the states $i^{*}$ and $f^{*}$ are the time-reversed relative to the states $i$ and $f$, and the momenta $p_a=m_av_a(a\equiv i,f.)$ for the reduced masses and the relative velocities of the two-body systems, we can construct the connection between the cross sections for the diagrams in Fig.~\ref{pair} and its inverse process
\begin{equation}\label{pde}
 \frac{\sigma_{f\overline{f}\rightarrow HH}}{\sigma_{HH\rightarrow f\overline{f}}}=\frac{1}{8}\frac{s-4m^2_H}{s-4m^2_f}.
\end{equation}
Note that here we have already considered the spin statistic weights and the symmetry factors for the processes. This relation holds true at all energies.

The r.h.s of Eq.~(\ref{pde}) would become the square of the Higgs boson velocity $v_{H}^{2}$ at non-relativistic energies, and we notice that the general theory of scattering requires that the cross section for the process $HH\rightarrow f\overline{f}$ is directly proportional to $v_{H}^{-1}$ in the low-energy limit and the other component of the cross section would be a constant \cite{Landau}. Using Eq.~(\ref{pde}), we can find that $\sigma_{f\overline{f}\rightarrow HH}$ should be in direct proportion to $v_{H}$ in the low-energy limit.

To the leading order in perturbation theory the cross section $\sigma^{p}_{e\overline{e}\rightarrow HH}=\lambda^{2}m_{e}^{2}v_{H}/(16 \pi m_{H}^{4})$ at the threshold. We here do not show the full expression which contains the extremely small contributions of the $t$- and $u$-channel diagrams, because it is cumbersome and no more instructive. It can be seen that the cross section has already got a factor $v_{H}$ in the expression. Consequently, we can conclude that the corrections from the diagram in Fig.~\ref{pair}d must be independent of the kinetic energies in the low-energy limit. The numeric simulations in Tab.~\ref{L-electron} show that the largest corrections to the cross sections cannot exceed ten percent.

\begin{table}[H]
\center
\caption{The corrections for the ``static'' Higgs boson pair production in the processes of electron-positron annihilation.}
\label{L-lepton}
\begin{tabular}{lc} \hline \hline

$~~~~~~~~~~~~~\sqrt{s}$(GeV)~~~~~~~~~~~~~~~~ &~~~~~~~~~~Correction(\%)~~~~~~~ \\\hline
~~~~~~~~~~~~~~~~251.001&9.85 \\ \hline\hline
\end{tabular}
\label{L-electron}
\end{table}

It should be noted that, as a matter of fact, a routine tree-level calculation can show that at the threshold $s$-channel diagram always absolutely dominates in the lepton pair annihilation processes. Hence, the results given above can also apply to muon and tau leptons.

\begin{figure}[h]
\center
 \subfigure[]{ \includegraphics[width=2.7cm]{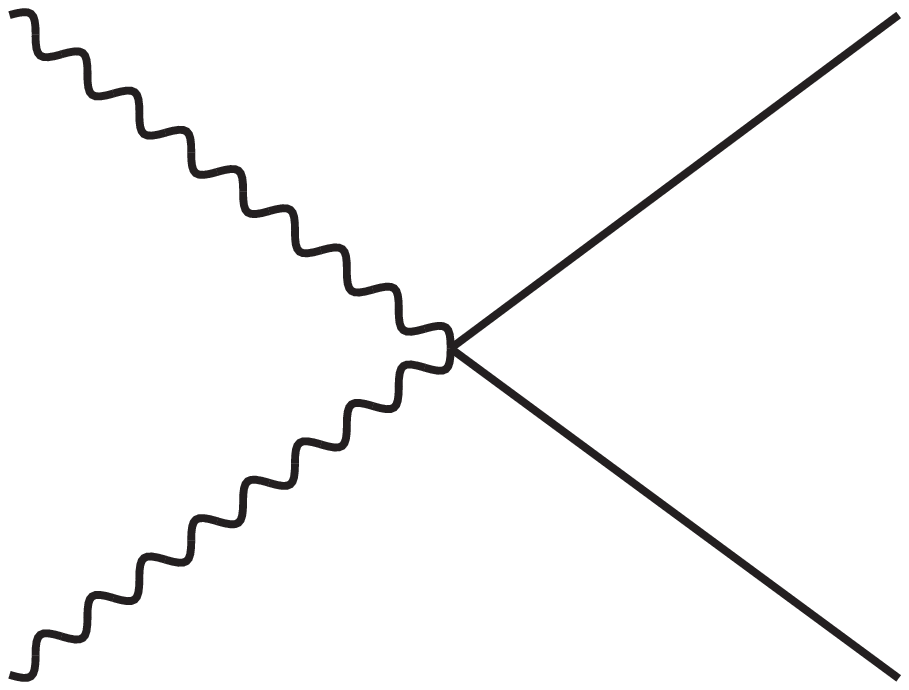}}~~~~~
 \subfigure[]{\includegraphics[width=2.7cm]{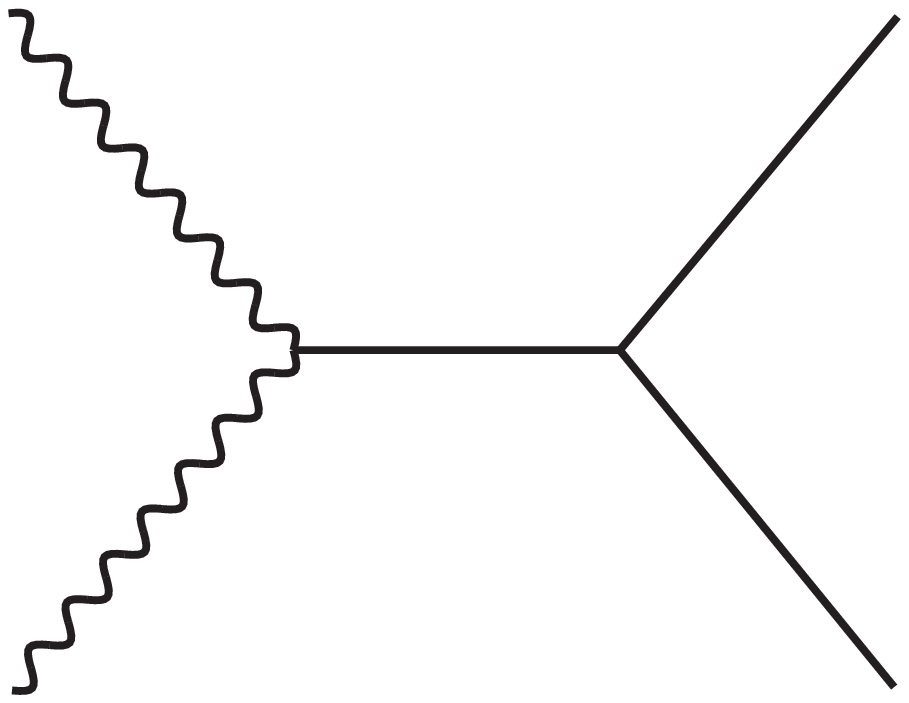}}\\
 \subfigure[]{\includegraphics[width=2.7cm]{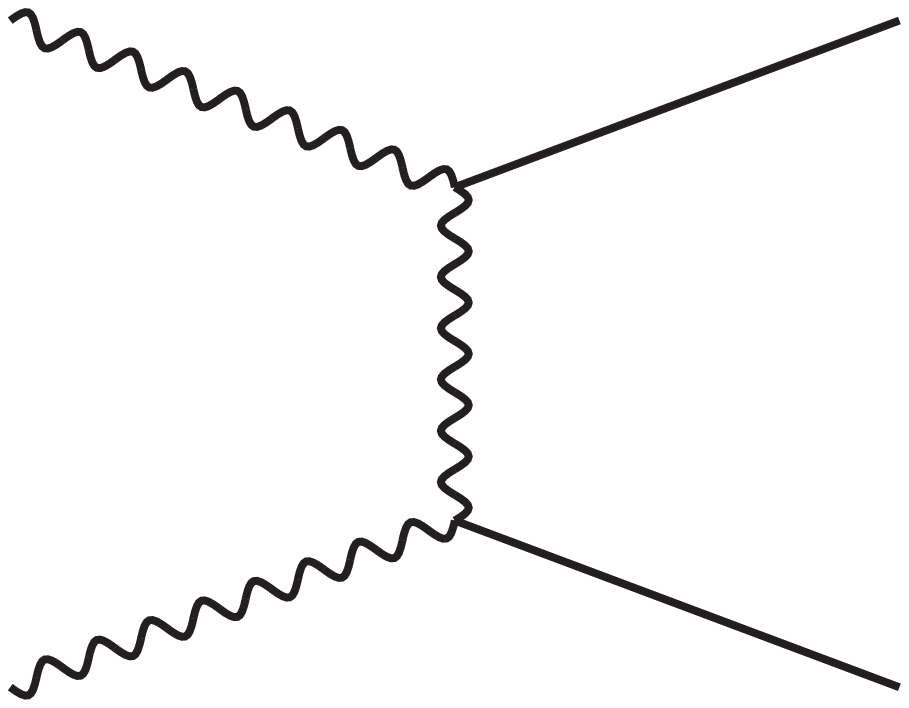}}~~~~~
 \subfigure[]{\includegraphics[width=2.7cm]{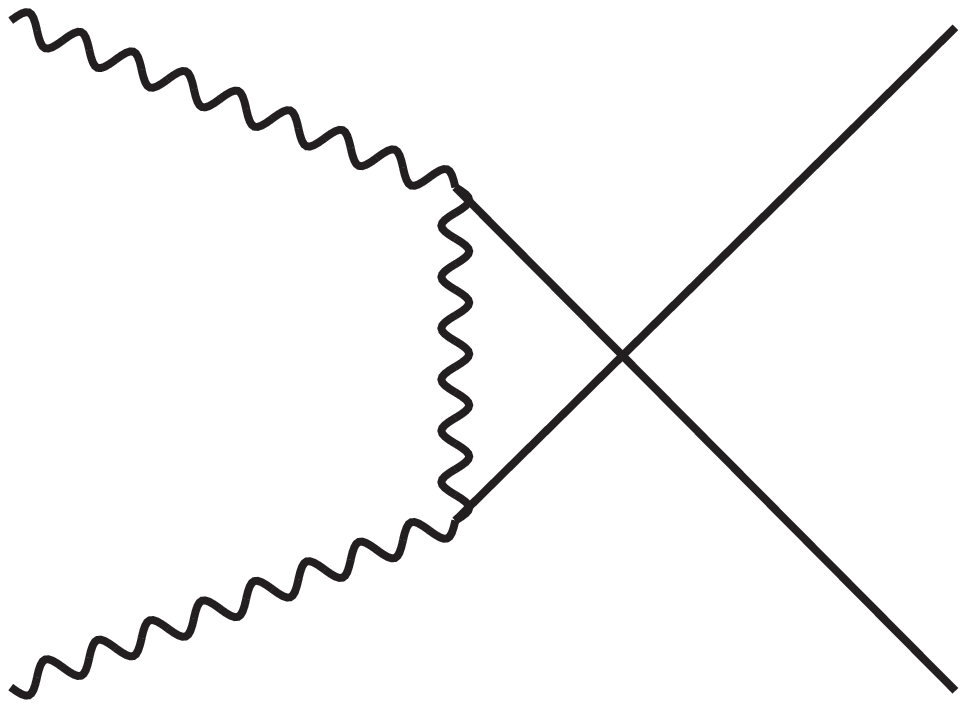}}
   \caption{Production of a Higgs boson pair in WW/ZZ fusion processes.}\label{fusion}
\end{figure}

The processes of $WW/ZZ$ fusion are important for the Higgs boson pair production in the GWS model, and now let us consider the final state interactions in these processes. The diagrams for the tree-level processes are shown in Fig.~\ref{fusion}.

The universal amplitude for the processes is
\begin{align}
\mathcal{M}=\mathcal{M}_{\text{quad}}+\mathcal{M}_{s}+\mathcal{M}_{t}+\mathcal{M}_{u},
\end{align}
where $\mathcal{M}_{\text{quad}}$ is the universal amplitude for the diagram in Fig.~\ref{fusion}a.

Notice that the contributions from the $t$- and $u$-channel diagrams can not be ignored in the processes. Thus, to find the corrections caused by the final state interactions, besides $\psi_f(0)$, we have to find $\psi_f(\bm r_t)$ and $\psi_f(\bm r_u)$ in the $\mathcal{M}_{t}$ and $\mathcal{M}_{t}$. This fact causes a serious problem for us to calculate the corrections caused by the effect of final state interactions. In principle, we cannot obtain the exact numeric simulations for the corrections, because it is impossible to get the $\bm r_t$ and $\bm r_u$ in the diagrams. However, there is a standard approach in quantum theory that can help us to offer the first theoretical predictions for the effect in WW/ZZ fusion processes. Notice that the energy of the non-relativistic Higgs boson $ E\simeq m_H$ and the uncertainty principle in the relativistic case is $\text{$\vartriangle${r}}\sim 1/E$~\cite{BLP}. It indicates that in the diagrams the distances around $1/m_{H}$ are important for the system of the created low-energy Higgs boson pair. To calculate the effect of final state interactions, we can assume that the $\bm r_t$ and $\bm r_u$ in the diagrams are $1/m_{H}$, and then the numeric simulations could approximate the strength of the corrections at the threshold.

Before doing the calculations, we should note that in the lepton pair annihilation processes we calculate the corrections of the squared modulus of the wave function at the origin, and then without any specific consideration there is only the s-wave of the $\psi^{0}_f$ that naturally contributes. However, here we shall not calculate the corrections to the s-wave at the origin in the $\mathcal{M}_{t}$ and $\mathcal{M}_{u}$. Thus we need to resolve the plane wave as $e^{i{k_fz}}=\Sigma^{\infty}_{l=0}(-i)^l(2l+1)P_l(\cos\theta)(r/k_f)^l(d/rdr)^l\sin{k_fr}/k_fr$ and then take account of the s-wave alone.

In the calculations we take $M_{Z}=91.188~\text{GeV}$ and $M_{W}=80.385$ GeV~\cite{PDG}, and we calculate the effect close to the threshold. The numeric simulations for the corrections in $ZZ$ fusion process are presented in Tab.~\ref{ZZ-fusion}. We find that the strength of the effect for $ZZ$ fusion is one order of magnitude weaker than that for lepton pair annihilation. The corrections in this process are small, but the effect might need to be considered for the high-precision experiments.

\begin{table}[h]
\center
\caption{The numeric simulations for the corrections in ZZ fusion process.}
\begin{tabular}{lc} \hline \hline

$~~~~~~~~~~~~~\sqrt{s}$(GeV)~~~~~~~~~~~~~~~ &~~~~~~~~~~Correction(\%)~~~~~~~~ \\\hline
~~~~~~~~~~~~~~~~251.01&0.853 \\
~~~~~~~~~~~~~~~~251.001&0.855 \\\hline\hline
\end{tabular}
\label{ZZ-fusion}
\end{table}

The numeric simulations for $WW$ fusion are shown in Tab.~\ref{WW-fusion}. We find that the strength of the effect for $WW$ fusion is even one order of magnitude weaker than that for $ZZ$ fusion. In general the effect in this process is negligible. Meanwhile, we notice that the numeric simulations for $WW$ fusion are sensitive to the distances of the interactions in the $t$- and $u$-channel diagrams, and in the future further studies on the accurate theoretical predictions for the corrections in $WW$ fusion process could be interesting.

\begin{table}[h]
\center
\caption{The numeric simulations for the corrections in WW fusion process.}
\begin{tabular}{lc} \hline \hline

$~~~~~~~~~~~~~~\sqrt{s}$(GeV)~~~~~~~~~~~~~~~ &~~~~~~~~~~Correction(\%)~~~~~~~ \\\hline
~~~~~~~~~~~~~~~~~251.01&0.082 \\
~~~~~~~~~~~~~~~~~251.001&0.084 \\ \hline\hline
\end{tabular}

\label{WW-fusion}
\end{table}

In summary, we have calculated the effect of final state interactions in three major processes of Higgs boson pair production for the first time, and we find that the corrections caused by the effect for these processes are markedly different. This paper could be useful for understanding and precisely determining the Higgs self-interactions at low energies.

\begin{acknowledgments}
The author is grateful to W. Shan for helpful discussions.
\end{acknowledgments}

\appendix
\section*{Appendix}

The ``cross'' term for the identical particles does not contribute to the non-relativistic potential. In this appendix, we shall give an elementary example in detail.

\begin{figure}[H]
\center
\subfigure[]{\includegraphics[width=2.8cm]{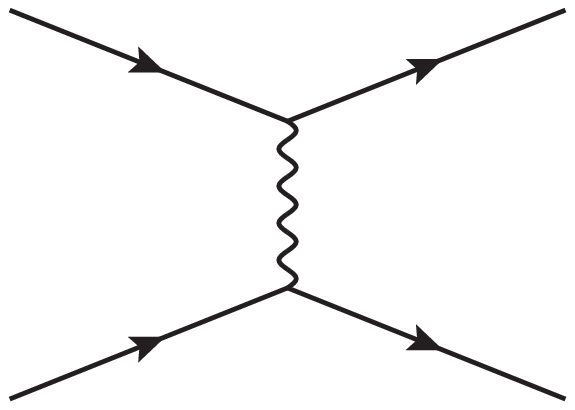}}~~~~
\subfigure[]{\includegraphics[width=2.8cm]{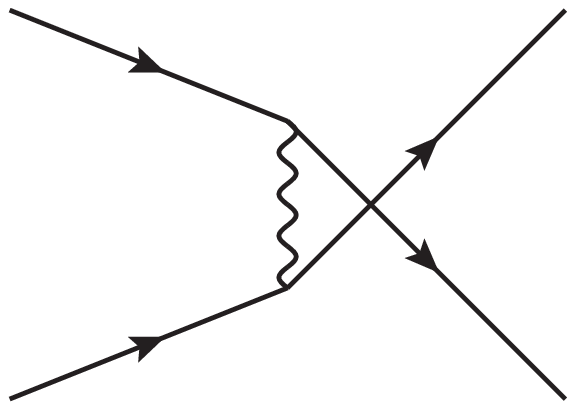}}
\caption{The scattering of two electrons in QED.}
\label{identical}
\end{figure}

Let us consider the scattering of two electrons in QED, see Fig.~\ref{identical}.

The cross section (M{/\kern-0.5em o}ller formula) is
\begin{align}\label{crosssection-qft}
 \frac{d\sigma}{d\Omega}_{\text{QED}}=&\frac{\alpha^2(2E^2-m^2)^2}{4E^2(E^2-m^2)^2} \nonumber \\ &\times[\frac{4}{\sin^4{\theta}}-\frac{3}{\sin^2{\theta}}+\frac{(E^2-m^2)^2}{(2E^2-m^2)^2}(1+\frac{4}{\sin^2\theta})],
\end{align}
where $m$ is the mass of electron.

In the non-relativistic domain, the cross section becomes
\begin{align}\label{NR-qft}
\frac{d\sigma}{d\Omega}_{\text{QED}}^{\text{NR}}=\frac{\alpha^2}{16m^2v^4}(\frac{1}{\sin^4\frac{1}{2}\theta}+\frac{1}{\cos^4\frac{1}{2}\theta}- \frac{1}{\sin^2\frac{1}{2}\theta\cos^2\frac{1}{2}\theta}),
\end{align}
where $v$ is the velocity of the electron.

It is well-known that the Coulomb potential $\alpha/r$ comes from the diagram in Fig.~\ref{identical}a for the distinguishable charged particles. Now let us consider the scattering of the identical electrons via the Coulomb potential $\alpha/r$ in non-relativistic quantum mechanics. The principle of identity requires that the asymptotic wave function must be
\begin{equation}\label{asymptotic-wave}
  \psi=e^{ikz}\pm e^{-ikz}+\frac{1}{r}e^{ikr}[f(\theta)\pm f(\pi-\theta)].
\end{equation}

It is elementary to find that the cross section for the process is
\begin{align}
\label{crosssection-qm}
\frac{d\sigma}{d\Omega}_{\text{QM}}=\frac{\alpha^2}{16m^2v^4}[&\frac{1}{\sin^4\frac{1}{2}\theta}+\frac{1}{\cos^4\frac{1}{2}\theta}\nonumber \\ &-\frac{1}{\sin^2\frac{1}{2}\theta\cos^2\frac{1}{2}\theta}\cos(\frac{\alpha}{2v}\ln\tan^2\frac{1}{2}\theta)].
\end{align}
In the perturbation region of the scattering process, Eq.~(\ref{crosssection-qm}) becomes the same as Eq.~(\ref{NR-qft}), see e.g.,~\cite{Landau}.

The example given above illustrates that the non-relativistic potential that we get from the ``distinguishable'' term will automatically take account for the contributions from the ``cross'' term for the identical particles.

\end{document}